\documentclass{article}
\usepackage{authblk}
\usepackage{graphicx}
\usepackage{cite}
\usepackage{url} 
\usepackage[margin=0.5in]{geometry}
\usepackage{footnote}
\usepackage{textcomp,gensymb}
\usepackage{amsmath}
\usepackage[flushleft]{threeparttable}
\usepackage{xcolor}

\begin{document}


\markboth{Aleksander Cianciara}{Simulation and Testing of a Linear Array of Modified Four-Square Feed Antennas for the Tianlai Cylindrical Radio Telescope}

\title{Simulation and Testing of a Linear Array of Modified Four-Square Feed Antennas for the Tianlai Cylindrical Radio Telescope}

\author{Aleksander J. Cianciara$^{\dagger\ast}$, Christopher J.  Anderson$^\ast$, Xuelei Chen$^{\S}$, Zhiping Chen$^{\ddagger}$, Jingchao Geng$^{\P}$, Jixia Li$^{\S}$, Chao Liu$^{\P}$, Tao Liu$^{\ddagger}$, Wing Lu$^\ast$, Jeffrey B. Peterson$^{\parallel}$, Huli Shi$^{\S}$, Catherine N. Steffel$^\ast$, Albert Stebbins$^{\bullet}$, Thomas Stucky$^\ast$, Shijie Sun$^{\S}$, Peter T. Timbie$^\ast$, Yougang Wang$^{\S}$, Fengquan Wu$^{\S}$, and Juyong Zhang$^{\ddagger}$}

\author[$\dagger$]{R. Campbell}
\author[$\star$]{M. Dane}
\author[$\ddag$]{J. Jones}

\affil[$\dagger$]{Department of Mathematics, Pennsylvania State University,Pittsburgh, Pennsylvania 13593}
\affil[$\star$]{Atmospheric Research Station,
Pala Lundi, Fiji}
\affil[$\ddag$]{Department of Philosophy, Freedman College,
Periwinkle, Colorado 84320}

\affil{
$^\ast$Department of Physics, University of Wisconsin - Madison, Madison, WI 53706, USA\\}

\affil{
$^{\S}$National Astronomical Observatories, Chinese Academy of Sciences, Beijing 100012, China}

\affil{
$^{\ddagger}$Hangzhou Dianzi University, Hangzhou, China}

\affil{
$^{\P}$The 54th Research Institute of the China Electronics Technology Group Corporation (CETC54), Shijiazhuang, China}

\affil{
$^{\parallel}$Department of Physics, Carnegie Mellon University, Pittsburgh, PA 15213, USA}

\affil{
$^{\bullet}$Fermi National Accelerator Laboratory, Batavia, IL 60510, USA}

\maketitle

\begin{center}
\begin{tabular}{l}
    Accepted for publication in Journal of Astronomical Instrumentation\\

\end{tabular}
\end{center}



\begin{abstract}
A wide bandwidth, dual polarized, modified four-square antenna is presented as a feed antenna for radio astronomical measurements. A linear array of these antennas is used as a line-feed for cylindrical reflectors for Tianlai, a radio interferometer designed for 21~cm intensity mapping. Simulations of the feed antenna beam patterns and scattering parameters are compared to experimental results at multiple frequencies across the 650 - 1420 MHz range.  Simulations of the beam patterns of the combined feed array/reflector are presented as well.
\end{abstract}


\section{Introduction}
Most of what we know about cosmology comes from measurements of the spatial and velocity distributions of baryonic and dark matter, i.e. the large scale structure (LSS) of the matter we can detect in the Universe.  Maps of LSS come from surveys of the distribution of galaxies, quasars and their absorption lines, and even cosmic microwave background (CMB) anisotropies and polarization patterns, which reveal LSS at the highest observable redshift. Since the quantum fluctuations that are thought to have seeded structure in the universe are stochastic, it is important to measure as large a volume of the LSS as possible in order to accurately determine the underlying distribution from which they were drawn.  The statistical properties of this distribution can tell us about the initial conditions of the universe (e.g. inflation), the properties of the matter it contains (e.g neutrino masses) and the overall geometry and cosmological dynamics (e.g. dark energy). Determination of these cosmological parameters from LSS statistical properties is largely limited by the size of the sample one obtains.  
Improving the precision of measurements of these parameters  therefore requires surveying large volumes of the universe. 

Galaxy redshifts can be measured at radio frequencies from the 21~cm hyperfine emission of atomic neutral hydrogen (HI). The 21~cm line is unique in cosmology because for $\lambda>21\,$cm it is the dominant astronomical line emission, 
so to a good approximation the wavelength of a spectral feature can be converted to a Doppler redshift without having to first identify the atomic transition. 
Unlike traditional optical galaxy surveys, 21~cm measurements do not rely on heavy elements from star formation, so they can in principle probe matter inhomogeneities from the cosmic dark ages before star formation, which account for the majority of the comoving volume of the observable universe. 

The 21~cm signal has been exploited to conduct galaxy redshift surveys out to $z\sim 0.1$ \cite{Zwaan2001, Martin2010}. 
Beyond this redshift, current radio telescopes do not have sufficient collecting area to make surveys using individual galaxies. 
The development of such an instrument would require a collecting area of order 1 square kilometer (the Square Kilometer Array), 100 times the collecting area of the Green Bank Telescope (GBT) or the Tianlai interferometer described here.  A radically different technique, intensity mapping, uses maps of 21~cm emission where individual galaxies are not resolved. Instead, it detects the combined emission from the many galaxies that occupy large ($1000~{\rm Mpc}^3$) voxels. The technique allows 100~m class telescopes such as Tianlai and the GBT, which only have angular resolution of several arc-minutes, to rapidly survey enormous comoving volumes at $z\sim1$ \cite{Abdalla2005,Peterson2006, Morales2008, Chang2008,Mao2008}.   
The overall promise of the intensity mapping technique has been studied by a number of authors \cite{Ansari2008,Seo2010,Ansari2012,Xu2015,Bull2015}.


The first measurements of the HI power spectrum, reported beginning in 2010, were made with the GBT using the intensity mapping technique at $z\sim 0.8$. HI maps made with the GBT were cross-correlated with maps of galaxy number counts from the DEEP2 and WiggleZ galaxy redshift surveys \cite{Chang2008, Chang2010, Masui2013, Switzer2013}.   The greatest challenge to these measurements is controlling systematic effects, particularly the removal of strong foreground radio emission \cite{Switzer2015}.  As demonstrated at the GBT, frequency-dependent cross-coupling from strong polarized sources into the intensity response of the telescope can introduce frequency-dependent contamination into the signal. Minimizing and measuring these spurious effects is critical. 

To survey large swaths of the sky with adequate signal to noise requires dedicated instrumentation. Both single dish and interferometric approaches are being developed.  BINGO\cite{Battye2013, Dickinson2014} will build an array of $\sim 50$ feed antennas for the focal plane of an off-axis reflector.  The 19-beam L-band focal plane for FAST could be used for an intensity-mapping survey as well\cite{Bigot2016}. A 7-beam array is under development for the 700-900~MHz band of the GBT\cite{Chang2016}. Most intensity-mapping survey instruments are interferometers and include cylindrical reflectors (Pittsburgh CRT\cite{Bandura2011},  CHIME\cite{Bandura2014}, the Tianlai cylinder array\cite{Chen2012}) as well as arrays of single dishes (Tianlai dish array and HIRAX\cite{Newburgh2016}). The  designs of these interferometers are related to those of wide-field arrays under development for studies of the Epoch of Reionization (EoR). EoR arrays operate at lower frequencies ($\sim 100$~MHz) to observe emission from HI at $z\sim 10$ and include LOFAR\cite{Yatawatta2013}, MWA\cite{Pober2016}, PAPER\cite{Ali2015}, and HERA\cite{Neben2016, Ewall2016}. 

\begin{figure}[htb]
\begin{center}
\includegraphics[width=5.05in]{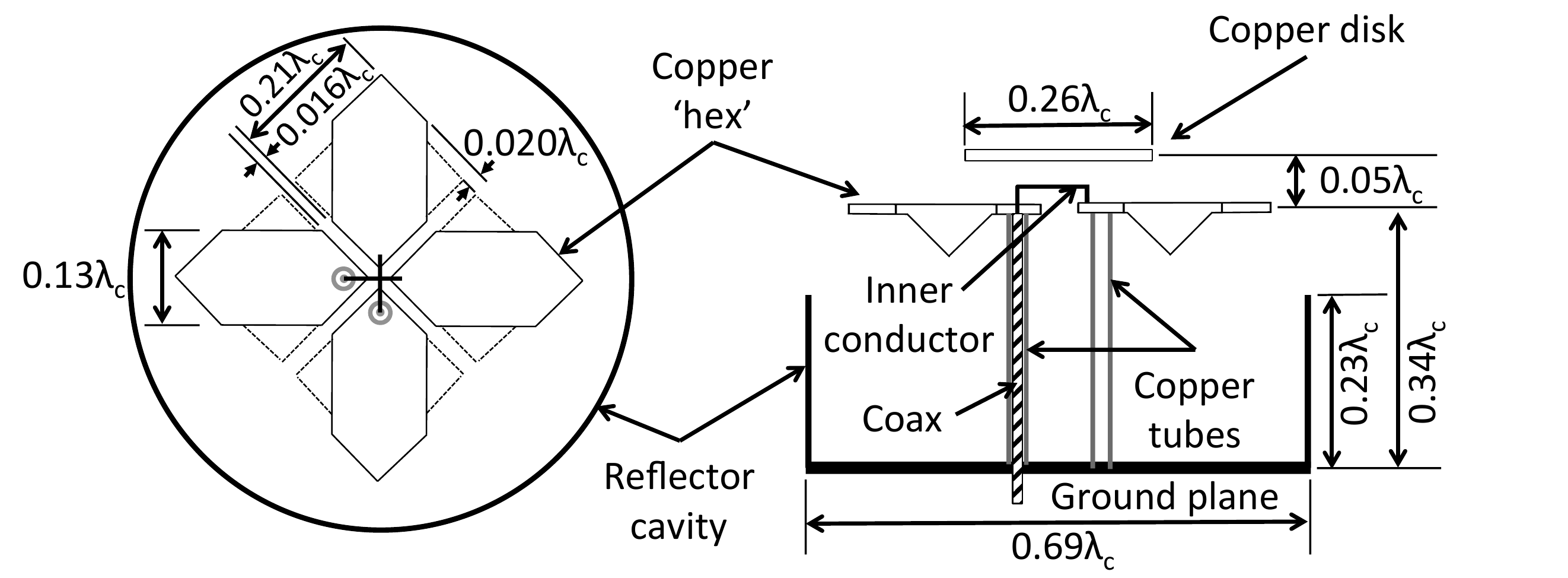}
\includegraphics[width=1.70in]{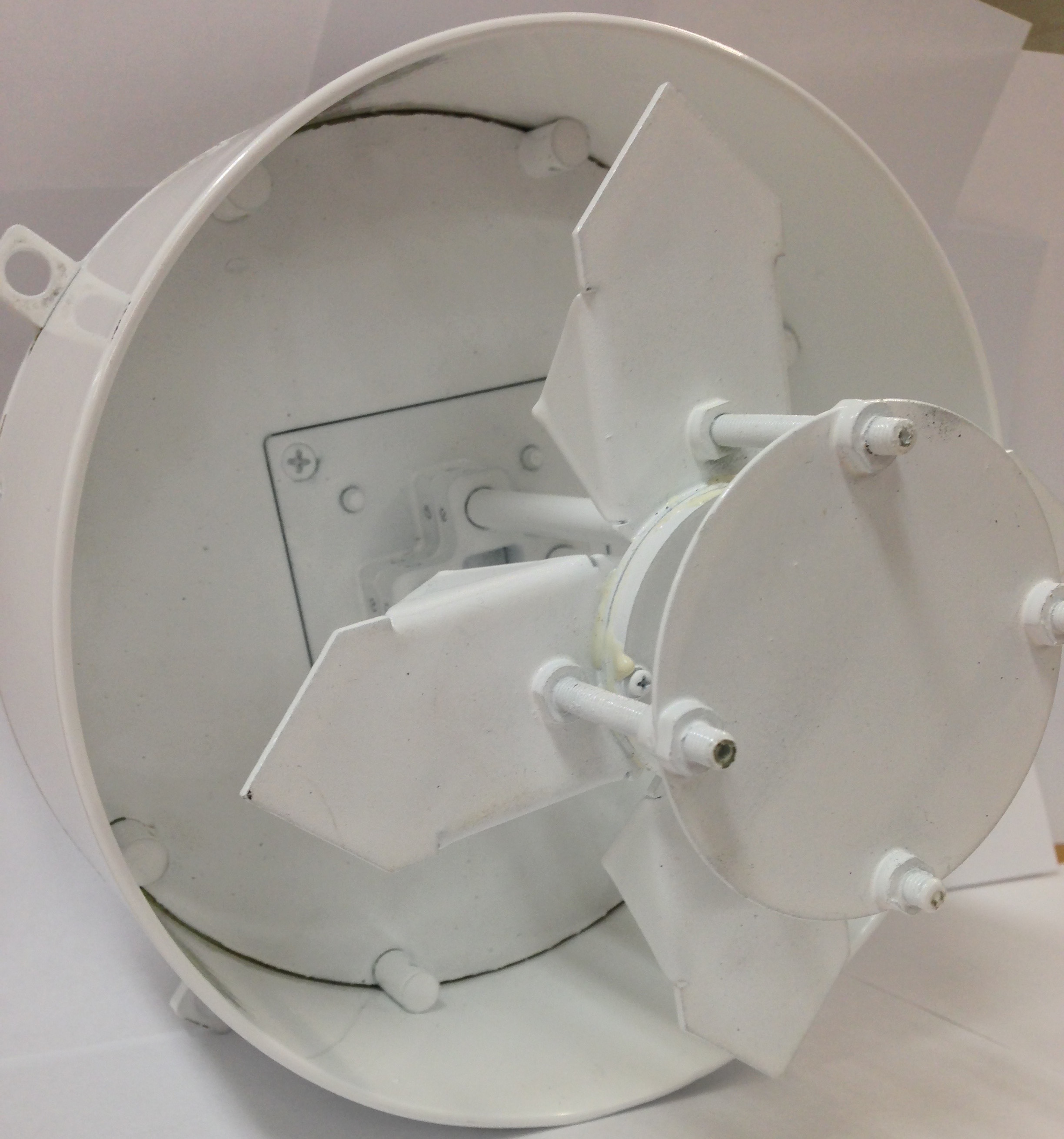}
\addtolength{\belowcaptionskip}{-7mm}
\caption{{\bf Left:} Schematic of feed.  Top view shows how triangular tabs are folded to convert a `four-square' to a `four-hex' shape. Dimensions are given in terms of $\lambda_c$, the wavelength at the center of the band, which is $\sim 29~$cm. The length of the dipoles, from `tip to tip' is $0.62\lambda_c$. {\bf Right:} Photo of feed.}
\label{fig:schematics}
\end{center}
\end{figure}

The Tianlai Pathfinders \cite{Chen2012} are located at a radio-quiet location ($44^\circ 10' 47''$~N $91^\circ 43' 36''$~E) in Hongliuxia,
Balikun County, Xinjiang Autonomous Region in northwest
China.  They consist of two independent interferometer arrays:  an array of 3, 15 m x 40 m cylindrical-parabolic reflector telescopes 
and an array of 16, 6~m diameter on-axis dish antennas. 
Both arrays will be tested in various configurations and, beginning in 2017, each array will begin a long-term drift-scan survey of the sky.

This paper discusses the Tianlai cylindrical array.  Parabolic cylinders have been used as reflectors for radio astronomy since about 1960 \cite{Ryle1960, Mills1963, Artyukh1967, Swarup1971,James2000, Bunton2002} as a means to create telescopes with large collecting areas at lower cost than either building a large single dish, phased arrays, or interferometers from multiple smaller antennas. One advantage of a cylindrical design is that it is relatively easy to achieve a large number of beams (corresponding to a large number of receivers) on the focal line. For an interferometer, the mapping speed is proportional to the total number of beams. Cylindrical telescopes currently in use include MOST\cite{Leung2007}, CHIME\cite{Bandura2014,Newburgh2014,Berger2016}, and Ooty\cite{Swarup1971}. Tianlai's cylindrical reflectors are located side-by-side with their long axes oriented along the N-S direction. The reflectors are fixed to stare at the zenith and execute a drift scan of the sky.  The cylindrical reflector focuses the beams in the E-W (drift scan) direction and produces a wide, unfocused, beam in the N-S direction.  Traditionally, a linear array of feed antennas placed along the focal line acts as a phased array to form beams along the cylinder axis. In the case of CHIME and Tianlai, rather than operating the feed antenna array as a phased array, the instruments are operated as interferometers; the signal from each feed antenna is correlated with signals from the other feed antennas along the feedline as well as feed antennas in different cylinders to form visibilities.


Below we present an analysis of the anticipated performance of the feed antennas for this cylindrical reflector interferometer. 
We give an overview of the design requirements of the instrument in Section \ref{overview}. In Section \ref{feed} we present the design, simulations, and measurements of the S parameters and beam patterns of a single, modified, four-square feed antenna. In Section \ref{cc_reflector}, we simulate the beam patterns of the cylindrical reflector fed by a single four-square feed antenna.
In Section \ref{array_reflector} we present simulations of the beam patterns and s-parameters of a single feed located within a focal line of feed antennas on a cylindrical reflector. A discussion of the implications of the feed design for measurements with Tianlai is presented in Section \ref{disc}.  We conclude in Section \ref{concl}. Future papers will discuss the Tianlai dish array, which uses a different type of four-square antenna optimized for feeding single dishes, and the calibration scheme.

\section{Design Overview} \label{overview}
A dedicated intensity mapping instrument has several design requirements that place significant constraints on the bandwidth and beam patterns of the feeds. While optimizing some of these design requirements is possible, trade-offs must be made. The design requirements given below are consistent with those assumed in recent forecasts for the Tianlai cylinder array \cite{Xu2015}.

A wide bandwidth is desirable because it allows maps to be made over a large range of redshifts, which increases the volume of the survey and enables the study of structure growth over cosmic time. In practice, it is difficult to achieve good impedance matching and desirable beam properties over extremely wide bandwidths. We attempt to achieve good properties over more than a factor of two in bandwidth, from 650~MHz (z = 1.2)
to 1420~MHz (z = 0), the 21~cm rest frequency.  This redshift range covers most of the epoch when dark energy began to dominate expansion, and the observations at low redshifts allow cross-correlation of the HI survey with low-redshift galaxy surveys. The Tianlai receivers have an instantaneous bandwidth of 100~MHz that can be tuned across this range of frequencies.  

The beam patterns of the feeds determine the field of view of the array.  For the cylindrical reflector design being considered, a wide field-of-view (in the non-focusing North-South direction) is critical because the telescope cannot be repointed, and so the North-South beam width determines the angular extent of the survey. Large survey volumes are required to reduce cosmic variance of measured parameters. Our goal is to map approximately $25\%$ of the sky.

Mapping speed depends on the feed patterns as well. Mapping speed is determined by the system noise temperature, $\mathrm{T_{sys}}$, which is the sum of three contributions: $\mathrm{T_{sys}=T_{rec} + T_{sky} + T_{spill}}$, where $\mathrm{T_{rec}}$ is the receiver noise temperature, $\mathrm{T_{sky}}$ is the antenna temperature of the sky, and $\mathrm{T_{spill}}$ comes from the feed antenna beam patterns spilling over the edge of the reflector and picking up thermal radiation from the warm ground. The mapping speed is very sensitive to the system temperature, scaling as $\mathrm{T_{sys}^{-2}}$, so it is critical to minimize $\mathrm{T_{sys}}$.  The value of $\mathrm{T_{rec}}$ is determined by the noise properties of the first stage low-noise amplifier: for Tianlai $\mathrm{T_{rec}\sim 25~K}$.  This value of $\mathrm{T_{rec}}$ will be treated as a given in this analysis, but we should note that any impedance mismatches between the antenna and amplifier will increase the effective value of $\mathrm{T_{rec}}$.  The sky temperature includes contributions from astrophysical backgrounds, $\mathrm{T_{bg}}$, and atmospheric emission, $\mathrm{T_{atmos}}$, so that $\mathrm{T_{sky}=T_{bg}+T_{atmos}}$, where  $\mathrm{T_{bg}=T_{CMB} + T_{Gal}}$, with $\mathrm{T_{CMB}=2.7~}$K and $\mathrm{T_{Gal}}$ is dominated by synchroton radiation from the Galaxy.  At high Galactic latitudes 
$\mathrm{T_{Gal}}$ ranges from $\approx 7.2 - 5.2~$K across the $700 - 800$~MHz band in which Tianlai will observe initially. The atmospheric noise contribution, $\mathrm{T_{atmos}}$, is approximately $3~$K across the bandwidth of this feed.  The sum of these contributions gives $\mathrm{T_{sky}}\approx 13~$K, subdominant to $\mathrm{T_{rec}}$. We require that the spill temperature decrease the mapping speed by no more than a factor of 2 (this corresponds to a $\mathrm{T_{spill}\approx 15~K}$) compared to a theoretical feed with $\mathrm{T_{spill}=0}$. To minimize $\mathrm{T_{spill}}$ requires that the beam pattern of each individual feed antenna be narrow enough to have very little spill beyond the width of the cylinder.  However, narrowing the beam too much will result in a narrow beam-width in the non-focusing direction of the cylinder and will limit the field of view of the survey. 

To first order the $\mathrm{T_{spill}}$ and field of view goals can be achieved with a feed that produces a symmetric, gaussian pattern with a 3~dB beamwidth of $\approx 45\degree$ in the North-South direction and that illuminates the edge of the cylinders, which have a f/D ratio of 0.32, with an edge taper of $\approx -10~$dB at the angle at which the edge of the reflector appears from the boresite of the feed, about $76\degree$,  The modified four-square feed described below approximates this behavior.


Finally, there are several particular challenges to designing the line feed for on-axis cylindrical reflectors like Tianlai's. First, the feeds must illuminate the cylinders with low edge taper while maintaining a small diameter to minimize scattering caused by blocking the beam in this on-axis design. To position `grating lobes' as far from the main beams of the array as possible, the feeds are placed close together. The feeds are nearly touching, with centers spaced $0.71~\lambda_c$ apart, where $\lambda_c \sim 29~$cm is the wavelength at the center of the band. However, close-spacing necessarily leads to cross-coupling between the feeds, which affects the beam patterns of the telescope and introduces correlated signals in the output of the correlator. 

Ultimately, the complex gain of the antennas and electronics must be measured in order to recover maps of the sky. This calibration process must be performed in real time using data from the sky. The goal of the present study is to provide realistic beam models to compare with those to be measured for the completed instrument.
Ideally, these beam patterns would be simple enough to be characterized by just a few parameters and be stable in time.

\section{`Four-hex' feed antenna} \label{feed}

\subsection{Design}
The four-square antenna has a long history \cite{Nealy1999, Stutzman2000, Buxton2001, Per2003, Suh2003, Bowman2006}, particularly as a feed antenna for cylindrical reflectors \cite{Leung2007, Leung2008, Bandura2014, Deng2014}. In the traditional four-square antenna the radiators are flat, square conductors, approximately $\lambda/2$ along the diagonal from `tip to tip,' and located $\approx \lambda/4$ in front of a ground plane. This radiator geometry provides a wider bandwidth than would linear dipoles. In our design (Fig. \ref{fig:schematics} - Left) the ground plane is a circular disk encompassed by a cylindrical rim of diameter $\sim \lambda/2$. The cylindrical `can' increases the gain and reduces cross-coupling between nearby feeds in an array. A description of a preliminary version of this antenna appears in Liu {\it et al.} (2014) \cite{Liu2014}. In the design reported here, two of the corners of each of the squares are folded toward the ground plane, creating a `four-hex' structure as seen from above. The folding is found to reduce cross-coupling between feeds, to reduce the coupling between the polarizations, and to improve the impedance match of the feed. In addition, a subreflector is located in front of the dipoles, similar to the design of a short-backfire antenna \cite{Kirov2009}. The subreflector creates a resonant structure which increases the gain. 
A standard quarter wave balun between the radiators and ground plane creates currents in the feed-line conductors that are equal in magnitude and opposite in phase, leading to a zero imbalance current, and hence does not affect the radiation pattern.  

\subsection{Single feed simulation and testing}
Simulations of scattering parameters and beam patterns were performed using CST Microwave Studio \cite{CST2016}, a 3D electromagnetic simulation software package. We used CST's time-domain transient solver and placed a waveguide port on the 50 Ohm coaxial line leeding to the excited dipole.  The accuracy of the solver was set to -35~dB, which means that the solver continued to calculate the field distribution and S-parameters until the electromagnetic field energy inside the structure had decayed to below -35~dB of the maximum energy inside the structure at any time. Likewise, the transient solver was run until the time signals at the ports had decayed to zero. For the measurements of $S_{11}$ and $S_{21}$ we used a HP VNA 8753ES. The calibration procedure included systematic error correction for directivity, source match, and frequency response.

\begin{figure}[!htb]
\centering
\includegraphics[width=3.4in]{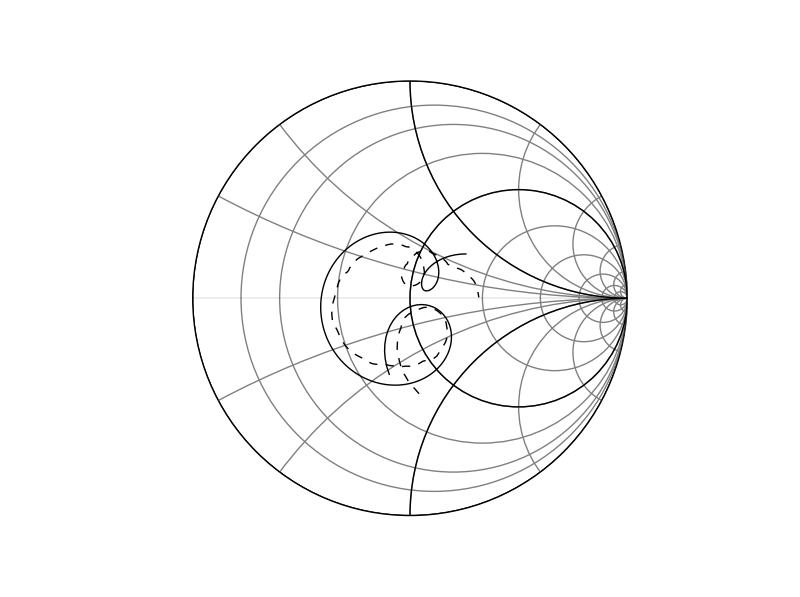}
\includegraphics[width=3.4in]{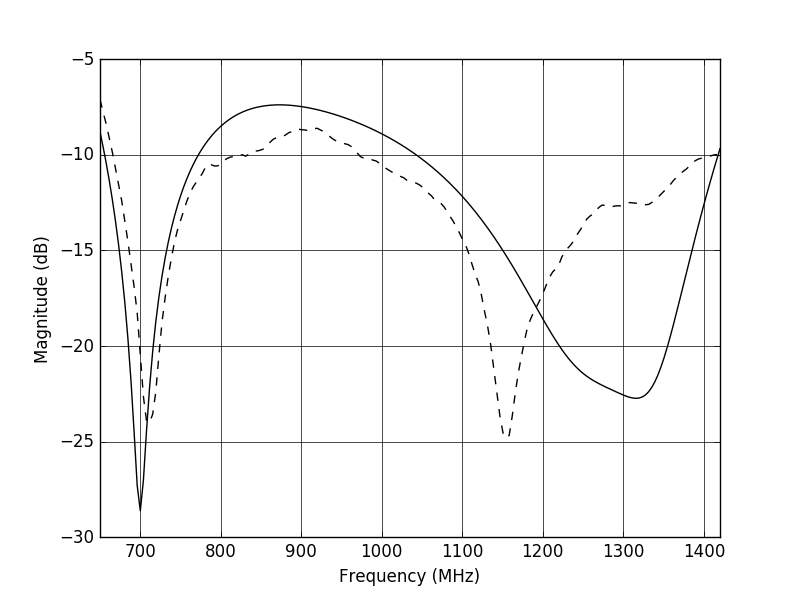}
\includegraphics[width=3.4in]{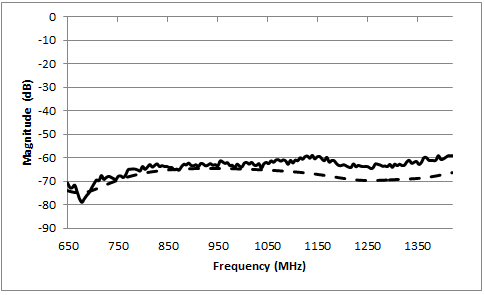}
\caption{Simulated and measured scattering parameters for a single feed antenna. {\bf Top Left:} The Smith chart for the $650 - 1420$~MHz band shows reasonable agreement between measured (dashed) and simulated (solid) values. (Curves start at 650~MHz at their lower left ends.) {\bf Top Right:} $S_{11}$ shows that reasonably good coupling is expected between the feed antenna and the following low noise amplifier, that is designed for a $50~\Omega$ impedance match over $\sim 650 - 1420$~MHz band. {\bf Bottom:} The measured and simulated values of the $S_{21}$ parameter show excellent isolation between the two polarizations of the feed across the entire band. The differences between measured and simulated parameters are probably due to a slight asymmetry in the fabricated antenna.}
\label{fig:Coffeecansparam} 
\end{figure}

Ports 1 and 2 were assigned to the two polarization outputs of the feed antennas. The modeled $S_{11}$-parameter (Fig. \ref{fig:Coffeecansparam}) shows a broad range below $\sim -7$~dB from $\sim 650-1420$~MHz and so provides an adequate impedance match to the Tianlai low noise amplifier, which is designed to match to a $50~\Omega$ load. The measured result is similar. The  $S_{21}$-parameter characterizes the cross-coupling between polarizations of the feed. The measured coupling is in reasonable agreement with the model. As shown below in Fig. \ref{fig:crosspol}, the cross polar response of the feed is far smaller than the cross polar response of the feed/reflector system. 

Beam patterns were also simulated using CST's time-domain transient solver with a far-field monitor for each frequency of interest. Patterns were measured in an anechoic chamber using a StarLab Multiprobe System manufactured by Satimo $\footnote{\url{www.mvg-world.com/en/products/field_product_family/antenna-measurement-2/starlab}}$. It is comprised of a rotation platform located in the center of a fixed circular arch that supports 16 equally spaced dual-polarization probe antennas.  The antenna under test undergoes an azimuthal rotation of 360 degrees that provides a full scan of its near-field pattern.  The far-field pattern is computed by a Fourier transform. The data acquisition as well as the gain calibration are performed by the standard propagation model (SPM).

Before the measurements are carried out, the gain of the system is determined so that the test antenna's absolute gain can be calculated in post-processing. The gain calibration is performed by running a standard measurement of a reference antenna (Satimo SH800 horn antenna) whose gain has already been measured in a reference measurement range. By comparing the reference antenna's gain to the measured antenna's gain, we get a calculated offset. This calculated offset is used as a reference level for transforming the near-field into a far-field. 

\begin{figure}[htb]
\begin{center}
\includegraphics[width=3.40in]{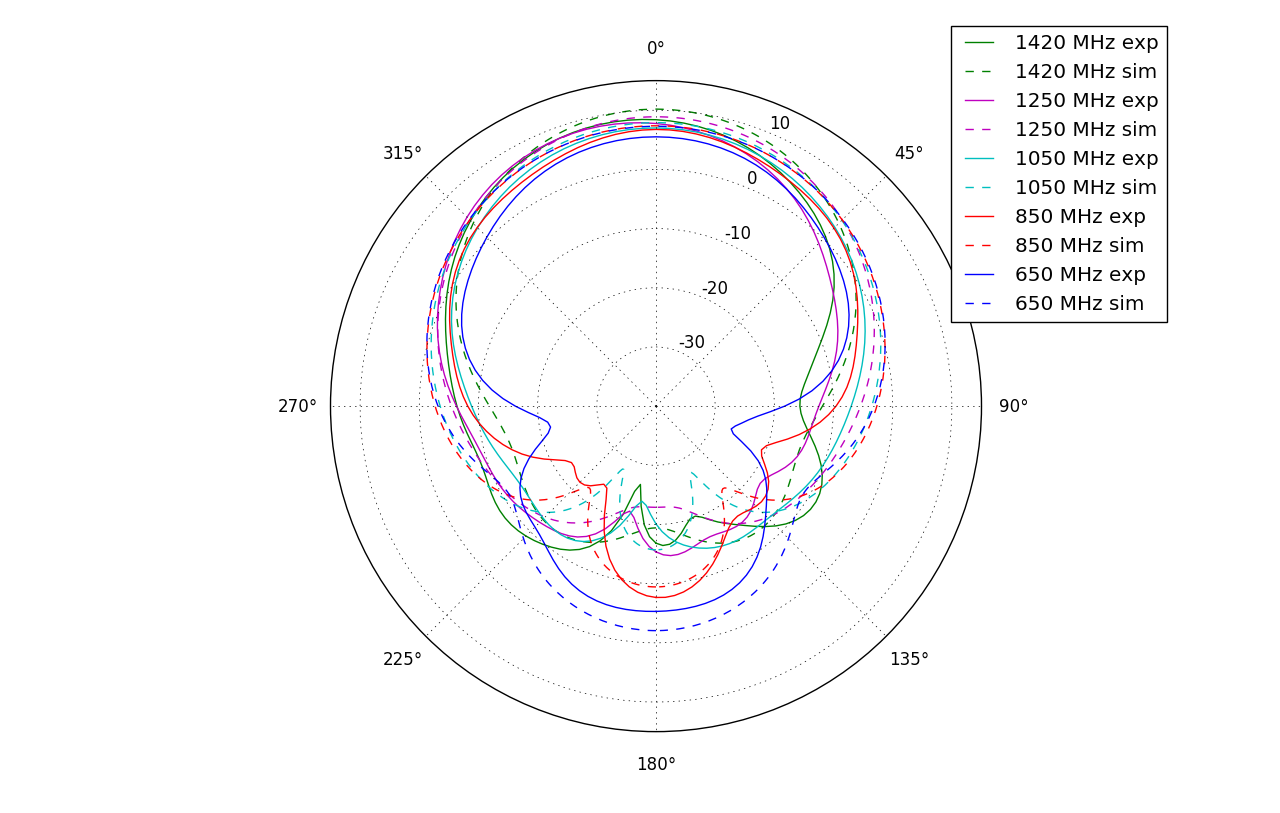}
\includegraphics[width=3.40in]{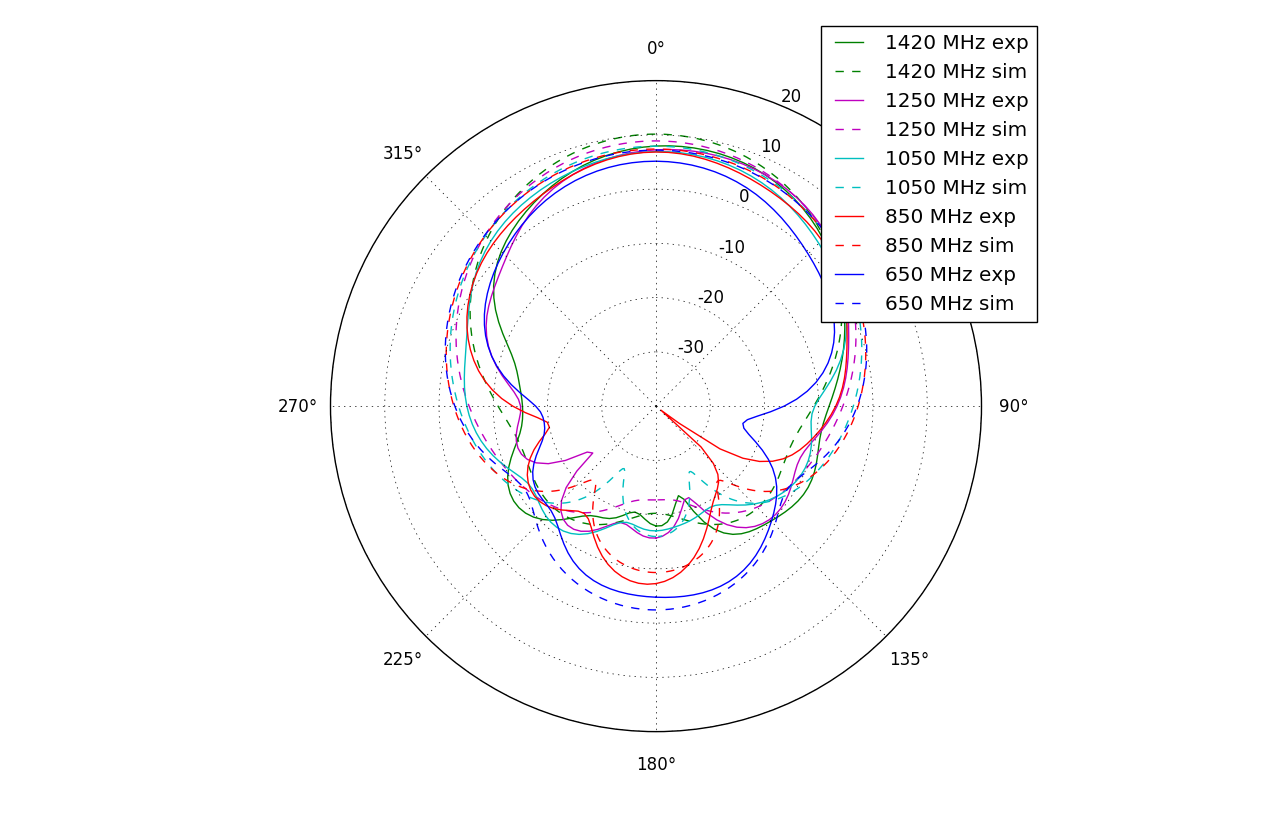}
\addtolength{\belowcaptionskip}{-7mm}
\caption{Measured and simulated beam patterns for the feed antenna. {\bf Left:} E-plane ($\phi = 0\degree$) and {\bf Right:} H-plane ($\phi = 90\degree$). Solid curves are the measured patterns;  dashed curves are simulated. }
\label{fig:1cc_0}
\end{center}
\end{figure}

\begin{figure}[htb]
\begin{center}
\includegraphics[width=3.40in]{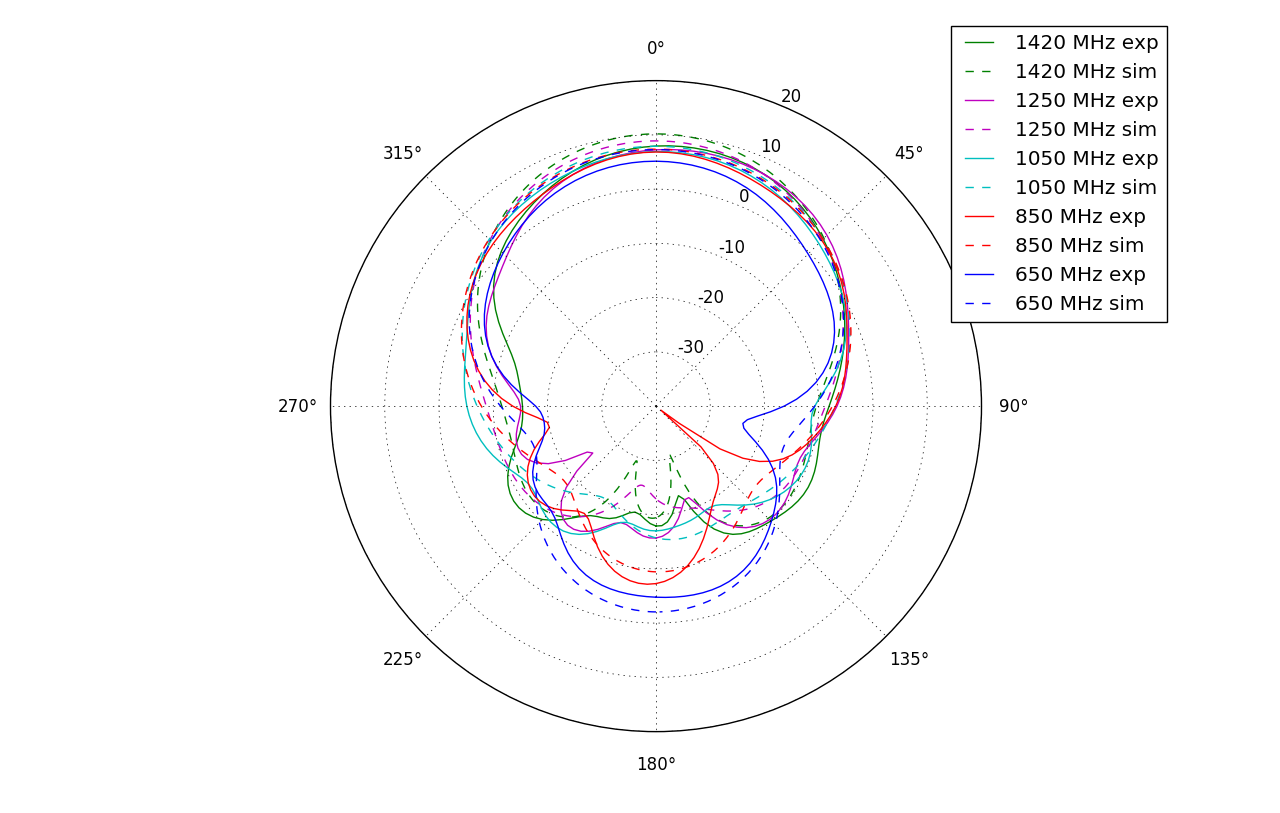}
\includegraphics[width=3.40in]{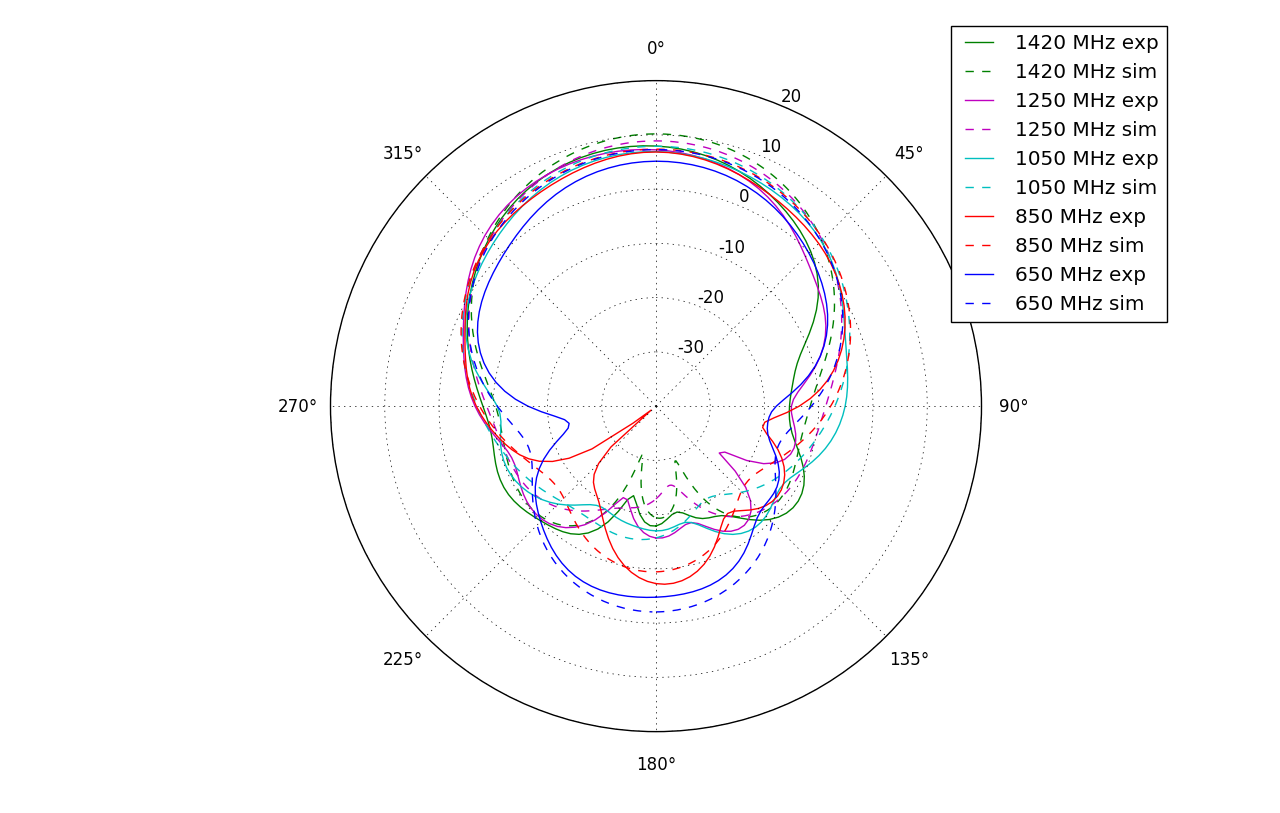}
\addtolength{\belowcaptionskip}{-7mm}
\caption{Measured and simulated beam patterns for the feed antenna. {\bf Left:} $\phi = 45\degree$ plane and {\bf Right:} $\phi = 135\degree$ plane. Solid curves are the measured patterns;  dashed curves are simulated.}
\label{fig:1cc_45}
\end{center}
\end{figure}

Figures \ref{fig:1cc_0} and \ref{fig:1cc_45} show simulated and measured beam patterns for the E-plane ($\phi = 0\degree$) and the H-plane ($\phi = 90\degree$) and the intermediate planes ($\phi = 45\degree$) and ($\phi = 135\degree$) for the feed antenna shown in Figure \ref{fig:schematics}. Here we define $\phi$ as the azimuthal angle between the plane of a simulated or measured beam pattern and the plane containing the axis of the dipole that is excited in a feed antenna.  

The measured patterns are in reasonable agreement with the simulated patterns. 
The edge taper specification ($\approx - 10~$dB at $82\degree$) is met by all of the simulated patterns and by the measured beam patterns at most frequencies between 650 and 1420~MHz.

\pagebreak

\section{Single Feed Antenna + Cylindrical reflector}
 \label{cc_reflector}
\begin{figure}[!htb]
\begin{center}
\includegraphics[width=3.35in]{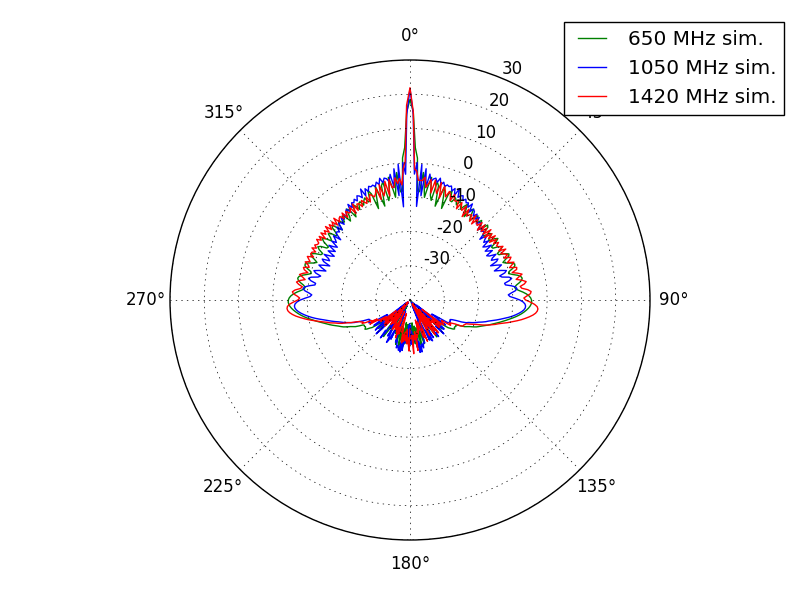}
\includegraphics[width=3.35in]{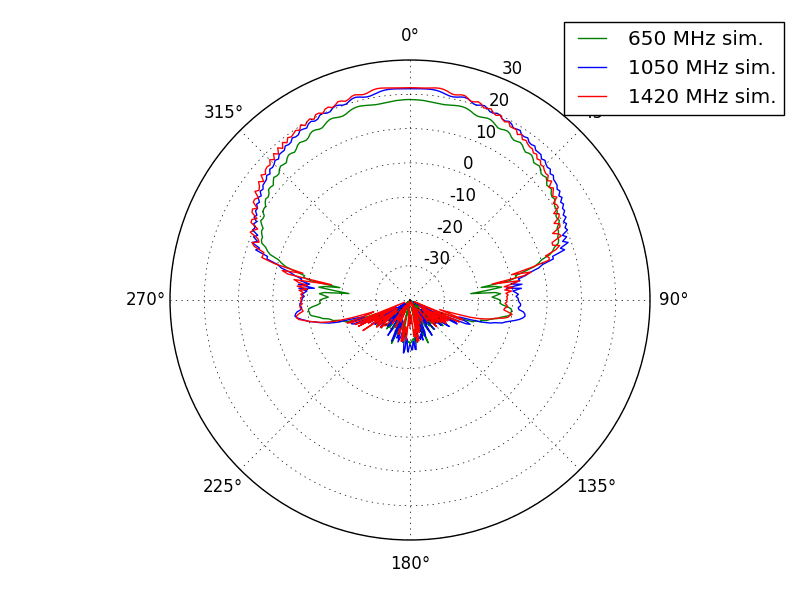}
\addtolength{\belowcaptionskip}{-1mm}
\caption{Simulated antenna patterns for the transverse dipole (perpendicular to the cylinder axis) of a single four-hex feed after reflection off the cylindrical reflector. {\bf Left:} $\phi = 90\degree$ (E-plane), FWHM at 1050~MHz = $1.2\degree$. {\bf Right:} $\phi = 0\degree$ (H-plane), Full width at half maximum (FWHM) at 1050~MHz = $61.4\degree$.}
\label{fig:1cc_reflector090T}
\end{center}
\end{figure}

\begin{figure}[!htb]
\begin{center}
\includegraphics[width=3.35in]{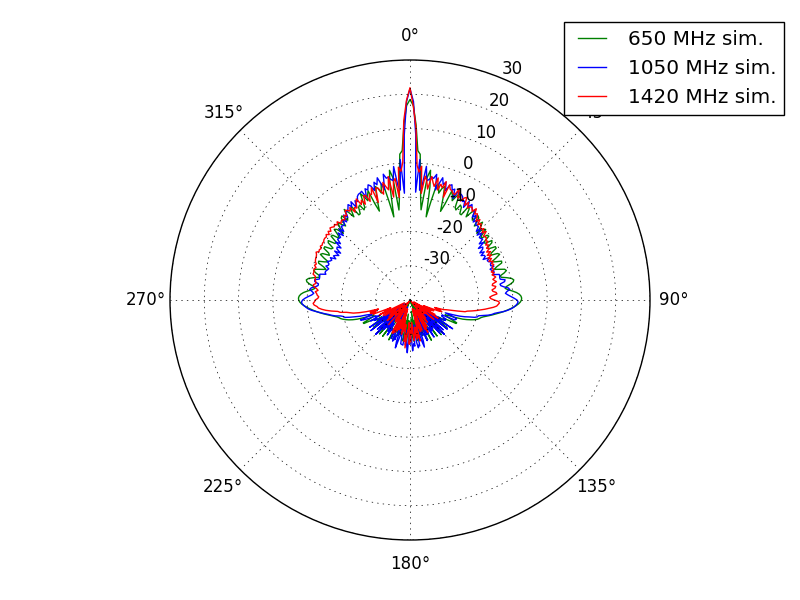}
\includegraphics[width=3.35in]{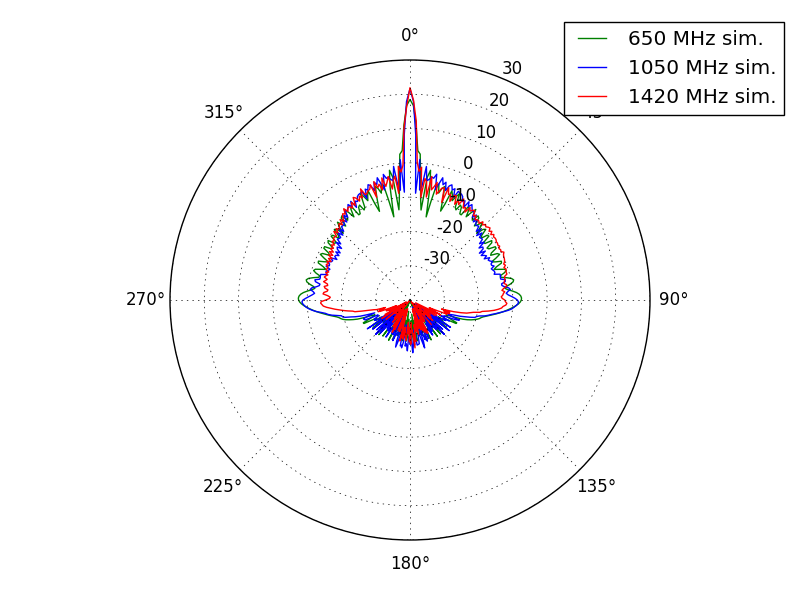}
\addtolength{\belowcaptionskip}{-1mm}
\caption{Simulated antenna patterns for the transverse dipole in a single four-hex feed after reflection off the cylindrical reflector {\bf Left:}  $\phi = 45\degree$, FWHM at 1050~MHz = $1.8\degree$. {\bf Right:}  $\phi = 135\degree$, FWHM at 1050~MHz = $1.8\degree$.}
\label{fig:1cc_reflector45135T}
\end{center}
\end{figure}

\begin{figure}[!htb]
\begin{center}
\includegraphics[width=3.35in]{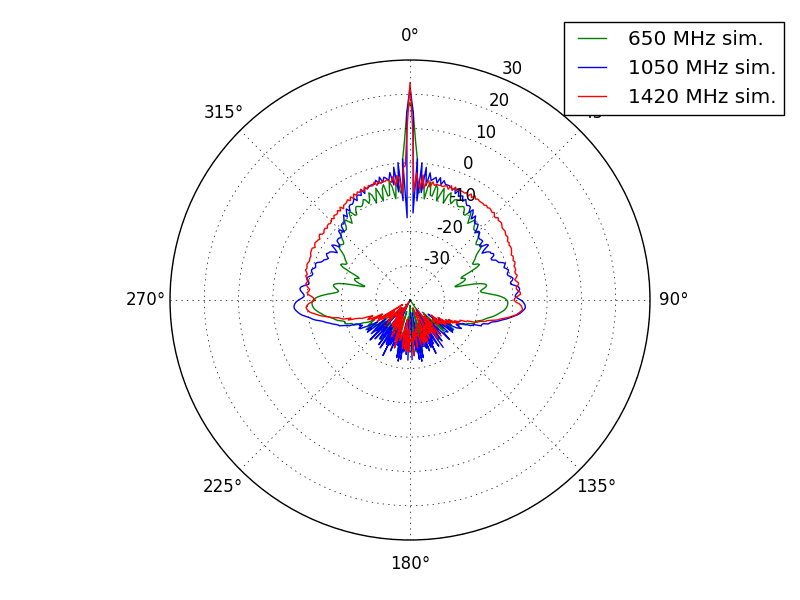}
\includegraphics[width=3.35in]{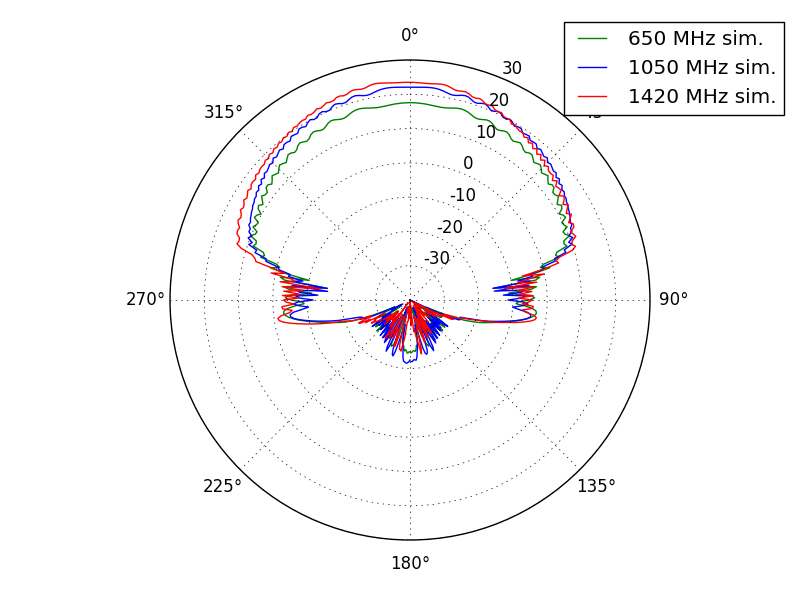}
\addtolength{\belowcaptionskip}{-1mm}
\caption{Simulated antenna patterns of the longitudinal dipole (parallel to the cylinder axis) of a single four-hex feed after reflection off the cylindrical reflector. {\bf Left:} $\phi = 90\degree$ (H-plane), FWHM at 1050~MHz = $1.2\degree$.  {\bf Right:} $\phi = 0\degree$ (E-plane), FWHM at 1050~MHz = 61.3$\degree$.}
\label{fig:1cc_reflector090L}
\end{center}
\end{figure}

\begin{figure}[!htb]
\begin{center}
\includegraphics[width=3.35in]{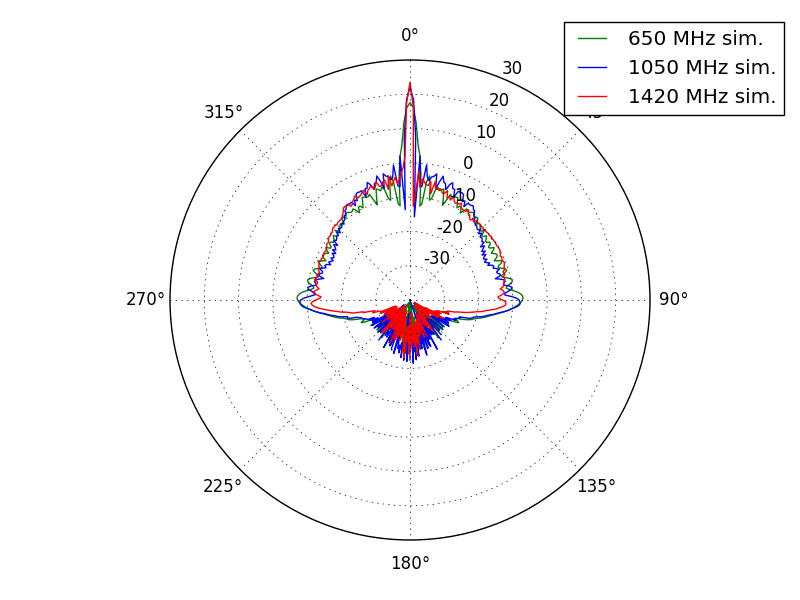}
\includegraphics[width=3.35in]{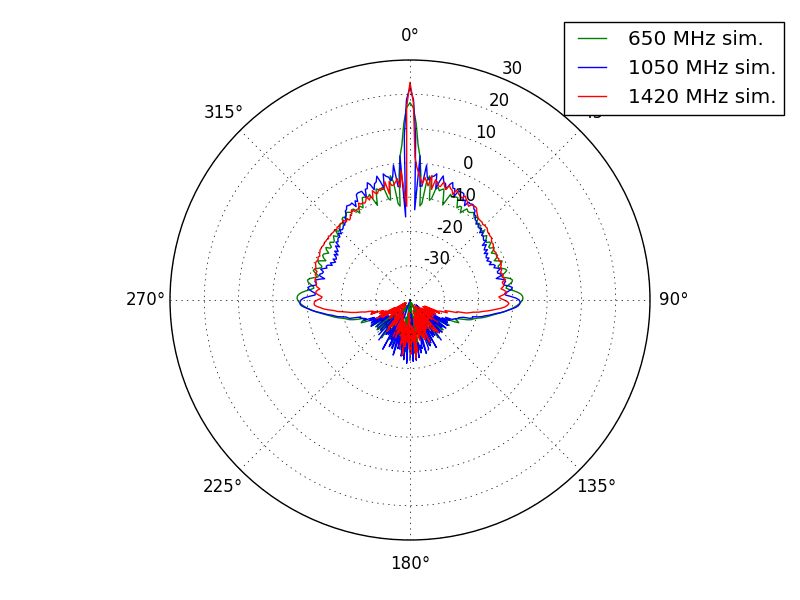}
\addtolength{\belowcaptionskip}{-1mm}
\caption{Simulated antenna patterns for the longitudinal dipole of a single four-hex feed after reflection off the cylindrical reflector. {\bf Left:} $\phi = 45\degree$, FWHM at 1050 MHz = $1.7\degree$. {\bf Right:} $\phi = 135\degree$, FWHM at 1050 MHz = $1.7\degree$.}
\label{fig:1cc_reflector0T}
\end{center}
\end{figure}

Next, we simulated the response of a single feed antenna with the cylindrical reflector. This case is studied to understand the effect of interactions between feed antennas that might occur when the feeds are assembled into a closely spaced linear array of feeds in Section \ref{array_reflector}. For the simulations including the cylindrical reflector, we define the $\phi = 0\degree$ direction to be along the length of the cylinder and the $\phi = 90\degree$ direction to be along the width of the cylinder.  For the as-built Tianlai cylinders, each feed has a dipole pointing along the length of the cylinder, which we call the longitudinal polarization, and an orthogonal dipole pointing along the width of the cylinder, which we refer to as the transverse polarization.  This convention continues throughout the following sections of the paper, and is illustrated in figure \ref{fig:schematic}. These simulations appear in Figures \ref{fig:1cc_reflector090T} - \ref{fig:1cc_reflector0T}.  
 

\section{Feed Antenna array + Cylindrical reflector}  \label{array_reflector}
Lastly, we simulate the beam patterns of a four-hex feed which is located at the center of a linear array centered on the line focus of a cylindrical reflector. The simulation includes the effects of focusing from the cylinder, cross-coupling between feeds, and blockage from the full array of 32 feeds. Here, as before, we used the time domain transient solver, with the accuracy again set to run until the electromagnetic field energy inside the structure had decayed to -35~dB of the maximum energy inside the structure at any time. The transient solver was run until the time signals at the ports had decayed to zero.
In the model, in addition to the array of feed antennas, we include the supporting structure for the feedline by inserting a rectangle of perfect electrical conductor running above the feeds on the focal line of the cylinder and with width equal to the diameter of the feeds. The cylindrical reflector is 15 meters wide in the focusing direction and 40 meters long. The f/D ratio is 0.32. In all simulations with the cylinder, the coordinate axes are set by the geometry of the cylinder (Figure \ref{fig:schematic}).

\begin{figure}[!htb]
\begin{center}
\includegraphics[width=3.0in]{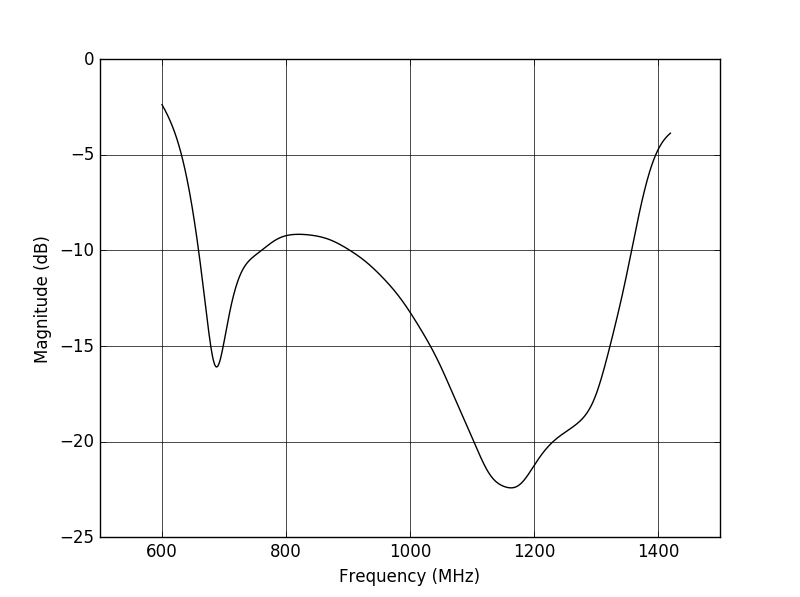} 
\includegraphics[width=3.0in]{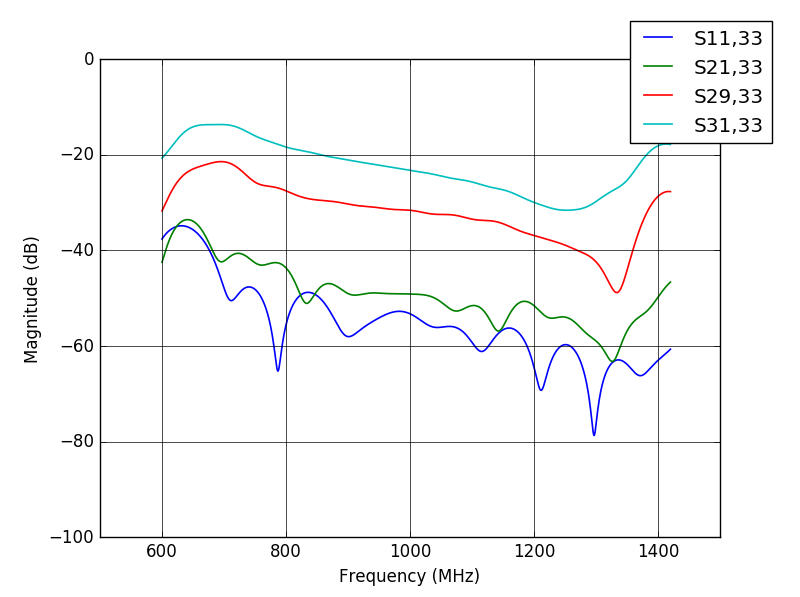} 
\includegraphics[width=3.0in]{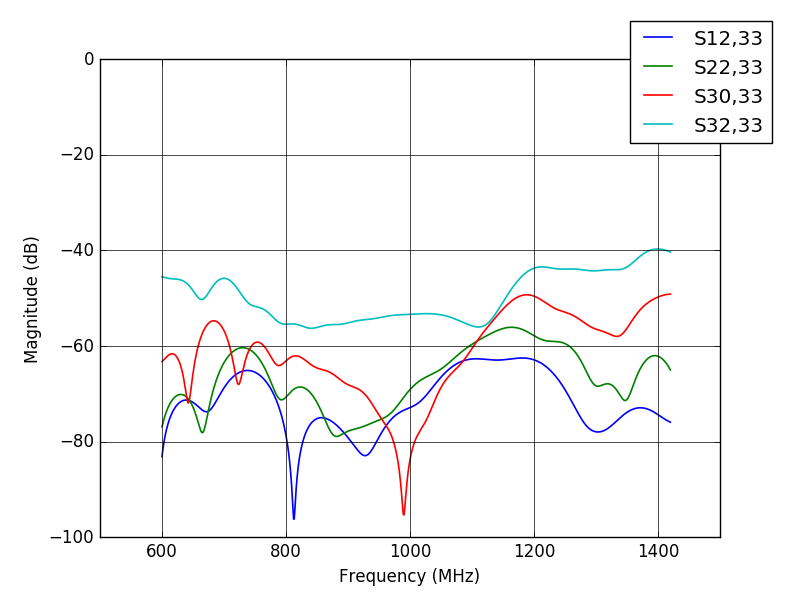}
\caption{S-parameters for feed antennas in an array at the line focus of a cylindrical reflector.  {\bf Top Left:} $S_{11}$-parameter for the transverse dipole on the 17th feed (next to the center) in an array of 32 feeds. In the notation used here this is the $S_{33,33}$ parameter (see text for port-numbering convention). Note the similarity to the case of a single feed without the reflector (Fig. \ref{fig:Coffeecansparam}). {\bf Top Left:} S-parameters for coupling between transverse dipole pairs.  {\bf Bottom:} S-parameters for coupling between transverse and longitudinal dipoles.}
\label{fig:cylindricalsparams}
\end{center}
\end{figure}

S-parameter simulations are shown in Fig. \ref{fig:cylindricalsparams} using the following port-labeling convention.  Odd-numbered ports correspond to dipoles oriented transverse to the long axis of the cylinder and are numbered 1, 3, ..., 63 from one end of the array to the other.  The even-numbered ports correspond to longitudinal dipoles and are numbered 2, 4, ..., 64.

Simulated beam patterns are shown for three frequencies in Figures 11-14. As expected, the cylinder focuses the beam in the plane perpendicular to the cylinder axis and does not focus in the plane parallel to the axis. The simulated FWHM in the focusing plane is consistent with expectations ($\sim 1.22 \lambda/D$). 

The contribution to the system temperature of ground spill from the array/reflector combination was computed from the simulated directivity for the 16th feed in the 32-element linear array (See Table \ref{tab:spill}.) The integral of the directivity that sees the ground was was divided by the integral of the directivity over all directions, and the result was multiplied by 290~K. In order to decrease mapping speed by no more than a factor of 2, we required a $\mathrm{T_{spill}} \approx 15~$K or less. This specification is met for both polarizations at 1420 MHz, the longitudinal polarization at 1050 MHz and 650 MHz, and approximately met for the rest. Note that the values of $\mathrm{T_{spill}}$ (from 15.5~K at 650 MHz to 7.1~K at 1420~MHz) are roughly consistent with the general trend of the frequency dependence of the edge taper achieved by feed antenna (Figs. \ref{fig:1cc_0} and \ref{fig:1cc_45}) at the cylinder edge, 76 degrees from the boresite of the feed, the simulated E and H plane patterns have an edge taper that improves with higher frequency, as expected, ranging from about -7~dB  at 650~MHz to about -17~dB at 1420~MHz. 

\begin{table}
\begin{center}
  \begin{tabular}{ | r | c | c | }
    \hline
    \multicolumn{3}{ |c| }{Ground spillover ($\mathrm{T_{spill}})$} \\
    \hline
	Frequency & Transverse dipole & Longitudinal dipole \\ \hline
    650~MHz & 15.5~K & 11.1~K \\ \hline
    1050~MHz & 15.1~K & 5.2~K \\ \hline
    1420~MHz & 7.1~K & 7.1~K \\ \hline
    \end{tabular}
    
\caption{\label{tab:spill} Computed values of $\mathrm{T_{spill}}$ for each dipole of the 16th feed in the 32-element array.}
\end{center}
\end{table}

\begin{figure}[!htb]
\begin{center}
\includegraphics[width=6in]{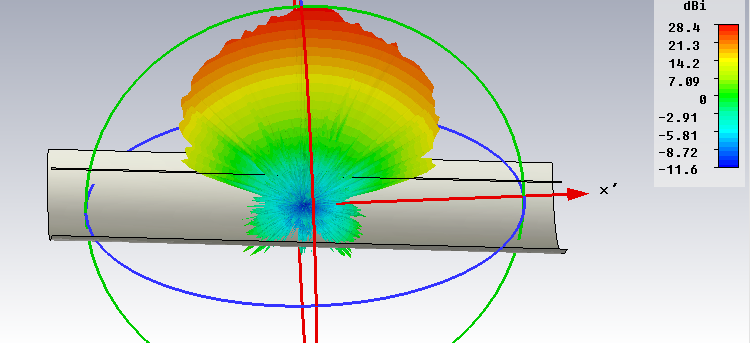}
\addtolength{\belowcaptionskip}{-1mm}
\caption{Schematic of the cylindrical reflector showing a 3D rendering of the directivity of the transverse dipole of the 16th four-hex feed in a linear array of 32 feeds. The angles $\phi$ and $\theta$ are defined as follows: $\phi = 0$ is oriented parallel to the long axis of the cylinder, and $\phi=90$ is transverse to the long axis. The $\theta=0$ direction is vertical.  The transverse and longitudinal dipoles are perpendicular and parallel to the length of the cylinder, respectively.}
\label{fig:schematic}
\end{center}
\end{figure}


\begin{figure}[!htb]
\begin{center}
\includegraphics[width=3.3in]{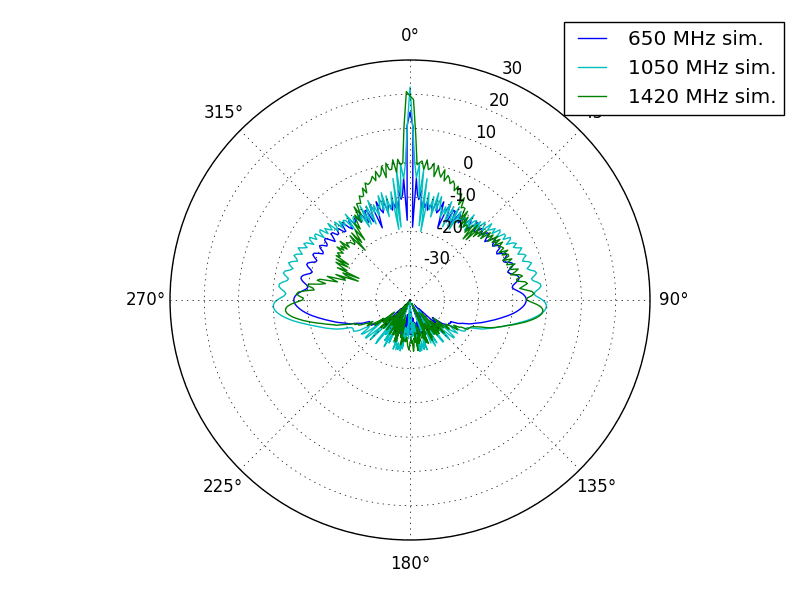} 
\includegraphics[width=3.3in]{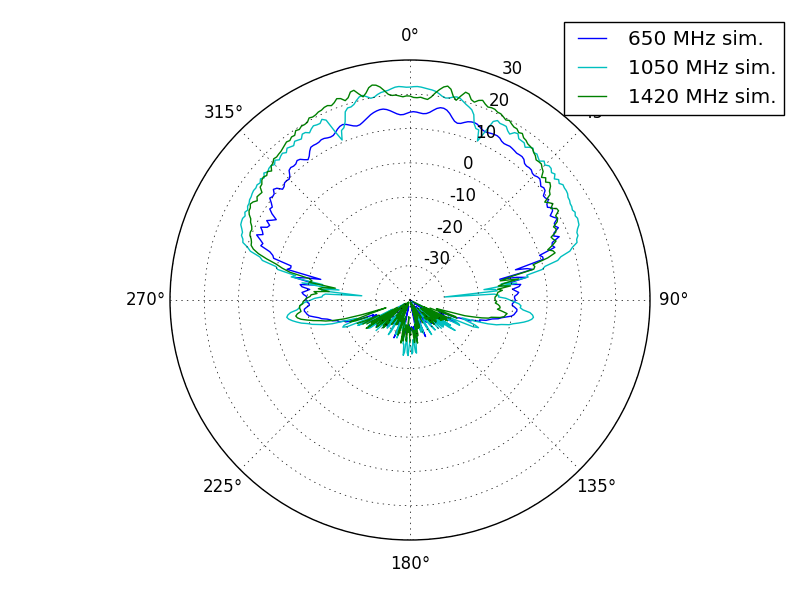}
\caption{Simulated antenna patterns for the transverse dipole of the 16th four-hex feed in a 32-element linear array, after reflection from the cylindrical reflector. {\bf Left:} $\phi = 90\degree$ (E-plane). The FWHM at 1050 MHz is 1.2$\degree$. {\bf Right:} $\phi = 0\degree$ (H-plane). The FWHM at 1050 MHz is 70$\degree$.}
\label{fig:rf0W}
\end{center}
\end{figure}

\begin{figure}[!htb]
\begin{center}
\includegraphics[width=3.3in]{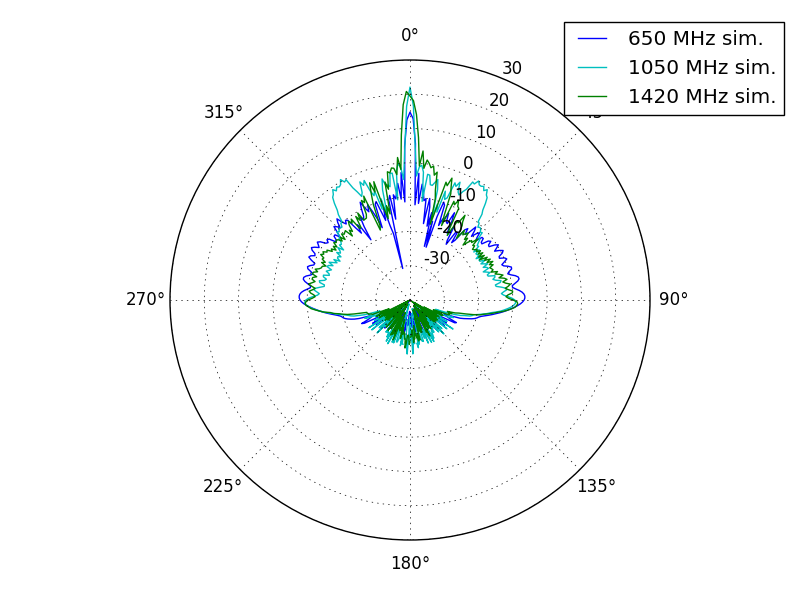}
\includegraphics[width=3.3in]{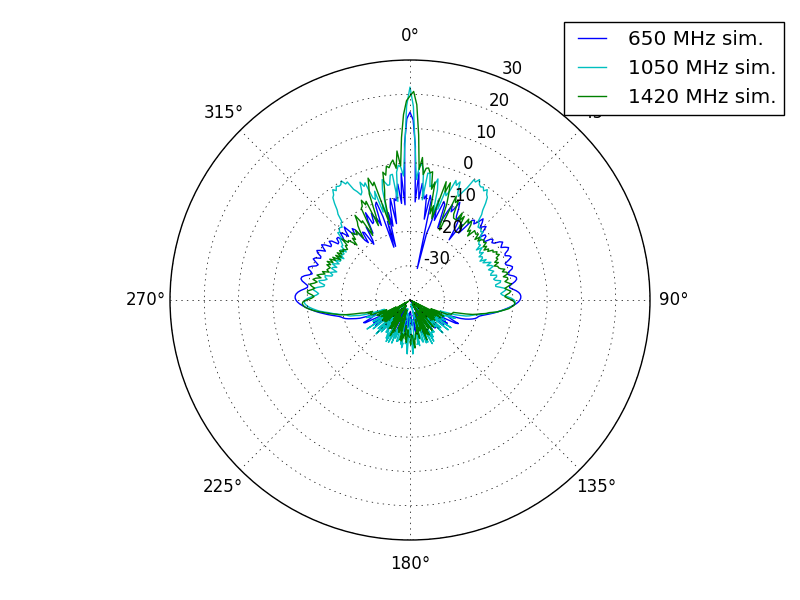}
\addtolength{\belowcaptionskip}{-7mm}
\caption{Simulated antenna patterns for transverse dipole of the 16th feed in a 32 element linear array, after reflection from the cylinder. {\bf Left:} $\phi= 45\degree$. The FWHM at 1050 MHz = 1.6$\degree$.  {\bf Right:} $\phi = 135\degree$.  The FWHM at 1050 MHz = 1.6$\degree$.}
\label{fig:rf45W}
\end{center}
\end{figure}

\begin{figure}[!htb]
\begin{center}
\includegraphics[width=3.3in]{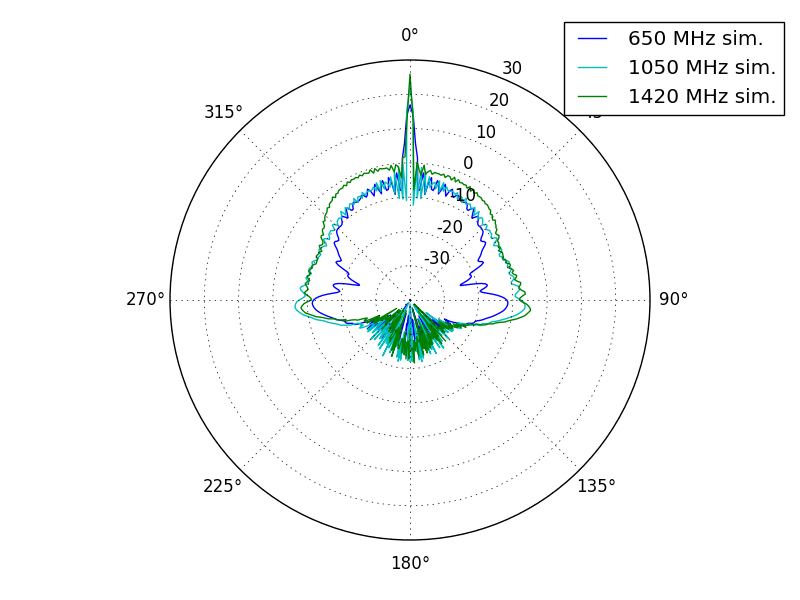}
\includegraphics[width=3.3in]{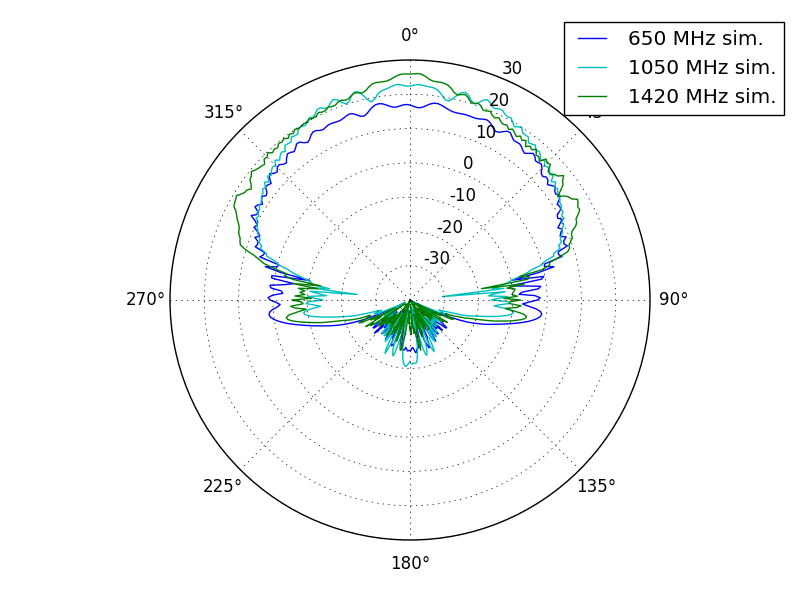}
\caption{Simulated antenna patterns for the longitudinal dipole of the 16th feed in a 32 element linear array, after reflection from the cylinder. {\bf Left:} $\phi = 90\degree$ (H-plane). The FWHM at 1050 MHz is 1.3$\degree$, {\bf Right:} $\phi = 0\degree$ (E-plane).  The FWHM at 1050 MHz is 56.1$\degree$.}
\label{fig:rf0L}
\end{center}
\end{figure}

\begin{figure}[!htb]
\begin{center}
\includegraphics[width=3.3in]{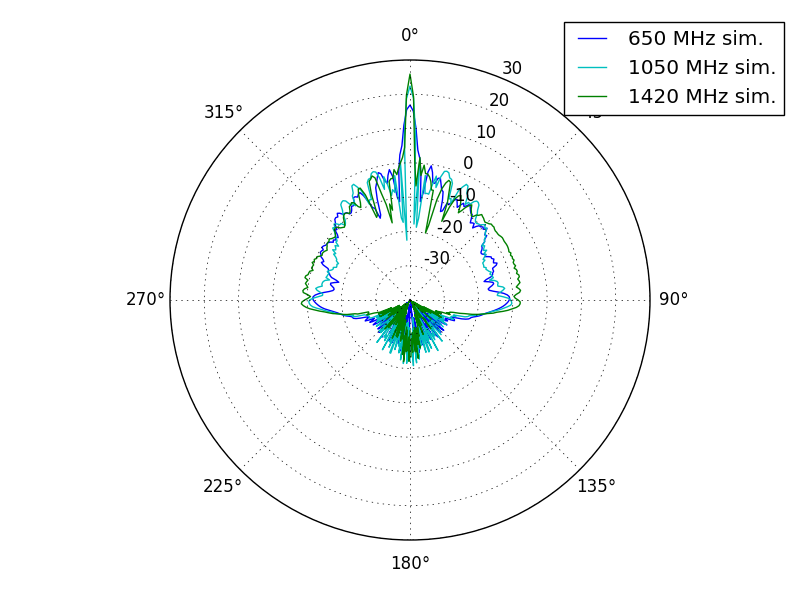}
\includegraphics[width=3.3in]{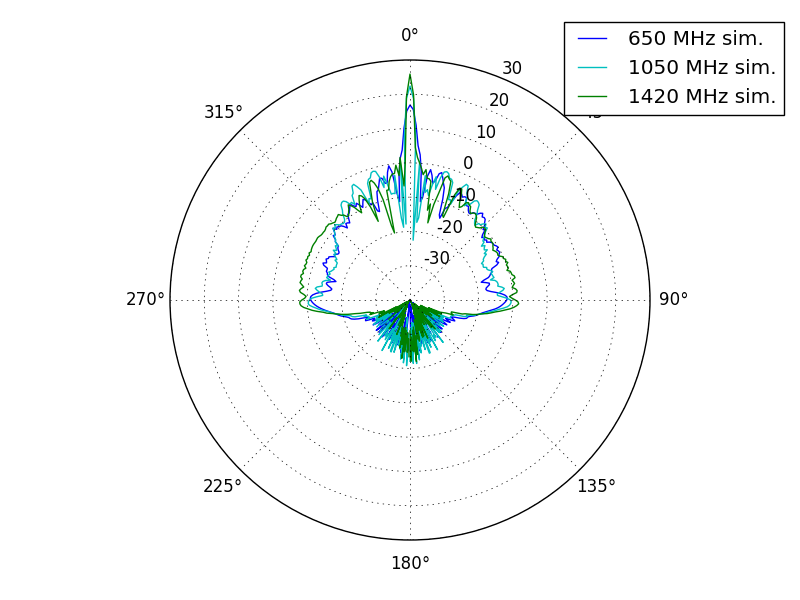}
\caption{Simulated antenna patterns for the longitudinal dipole of the 16th feed in a 32 element linear array, after reflection from the cylinder.  {\bf Left:} $\phi = 45\degree$.  The FWHM at 1050 MHz is 1.8$\degree$. {\bf Right:} $\phi = 135\degree$.  The FWHM at 1050 MHz = 1.8$\degree$. }
\label{fig:rf135L} 
\end{center}
\end{figure}

\begin{figure}[!htb]
\begin{center}
\includegraphics[width=3.3in]{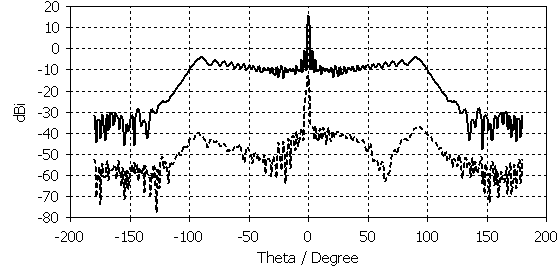}
\includegraphics[width=3.3in]{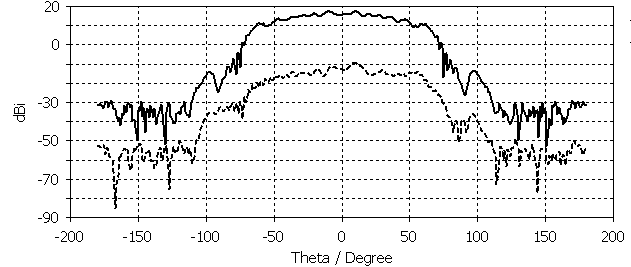}
\includegraphics[width=3.3in]{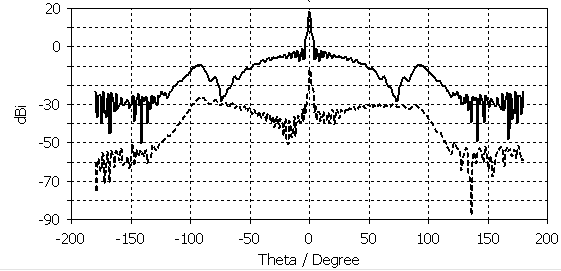}
\includegraphics[width=3.3in]{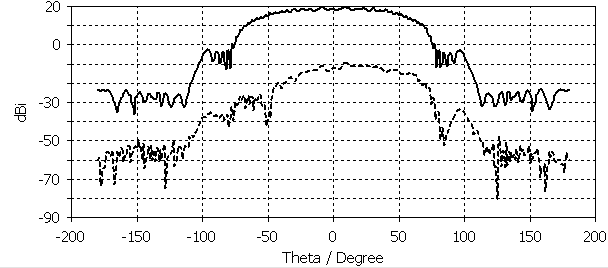}
\caption{Simulated co-polar (solid) and cross-polar (dashed) antenna patterns for the 16th feed in a 32 element linear array, after reflection from the cylinder, at 650 MHz.  {\bf Top Left:} Longitudinal dipole, $\phi = 90\degree$ (H-plane).  {\bf Top Right:} Longitudinal dipole, $\phi = 0\degree$ (E-plane). {\bf Bottom Left:} Transverse dipole, $\phi = 90\degree$ (E-plane). {\bf Bottom Right:} Transverse dipole, $\phi = 0\degree$ (H-plane).}
\label{fig:crosspol} 
\end{center}
\end{figure}

The co-polar and cross-polar response for a feed near the center of the line feed is shown in Figure \ref{fig:crosspol}. The cross polar pattern is determined using the Ludwig - 3 definition of cross polarization.  At most values of $\theta$ the cross polar response is 20~dB or more below the co-polar response.

\section{Discussion}  \label{disc}

A comparison of the beam pattern for a feed near the center of the full feedline and reflector (Figures \ref{fig:schematic} - \ref{fig:rf135L}) to the case of a single feed and reflector (Figures
\ref{fig:1cc_reflector090T} -
\ref{fig:1cc_reflector0T}), shows that the overall sidelobe level in the full feedline case is increased slightly. The gain of the main beam remains about the same.  Presumably, the increase in sidelobe level occurs because of beam blockage and scattering by the feedline.  The cross-sections of the beam patterns in the non-focusing direction show more structure than do the single-feed/reflector combination and both show more structure than the feed by itself.  Calibration of the full telescope with high angular resolution will be required to resolve these features. 
The groundspill temperature, $\mathrm{T_{spill}})$, of the simulated antenna is less that 15.5~K across our band, very close to our requirement of 15~K.  The 50~K system temperature assumed in the forecasts of \cite{Xu2015} for Tianlai is consistent with this analysis.  Similarly, the 3~dB beamwidth traces out a swath of the sky greater than or equal to 10,000 square degrees at all simulated frequencies, consistent with the estimate of \cite{Xu2015}.  Figure \ref{fig:cylindricalsparams} shows that the cross coupling between feeds is below -15~dB at all frequencies, even for neighboring parallel polarizations, and drops off rapidly at larger spacings.  This coupling can add at most a few Kelvin to the system temperature of the shortest baselines, but again the forecasts of \cite{Xu2015} will be basically unchanged. Though it is true that the beam patterns are not simple enough to be described by just a few parameters, as hoped, such a simple model could not be accurate to high precision anyway.  The large dynamic range of foregrounds compared to signal for any 21~cm intensity mapping experiment means that an accurate and detailed beam model is essential.  In the transit interferometer m-mode analysis developed in \cite{shaw2014all}, for example, a full spherical harmonic expansion of the beam is needed to disentangle the sky from the properties of the beam.  Successful foreground removal will depend on the accuracy of the beam model, built through simulation and careful calibration.  

The polarization properties of the beam pattern are another important systematic effect that can cause foregrounds to become problematic.  Foreground removal techniques generally rely on the inherent smoothness of the foreground frequency structure compared to the clumpy HI signal.  However, polarization leakage can impart a complicated frequency structure on inherently smooth polarized foregrounds.   This can happen as follows: consider a linearly polarized source of power $P$ at an angle $\theta$ with respect to the co-polarization axis of the transverse dipole.  The power detected by the transverse and longitudinal dipoles will be
\begin{equation}\label{eq:t}
P_t = C_tP\cos^2(\theta) + X_tP\sin^2(\theta) + 2\sqrt{C_tX_t}\sin(\theta)\cos(\theta)\cos(\phi_t)
\end{equation}
\begin{equation}\label{eq:l}
P_l = C_lP\sin^2(\theta) + X_lP\cos^2(\theta) + 2\sqrt{C_lX_l}\sin(\theta)\cos(\theta)\cos(\phi_l)
\end{equation}
The Stokes $I$ intensity due to this source is the sum of powers from the two orthogonal dipoles.  
\begin{equation}\label{eq:I}
I = P_t + P_l
\end{equation}
If the intensity $I$ changes as a function of polarization angle, this is known as polarization leakage of linearly polarized intensity into the Stokes $I$ parameter.  This leakage is dangerous in 21~cm experiments because (1) polarized synchrotron foregrounds tend to be Faraday rotated, so that their polarization angle changes as a function of frequency and (2) the leakage response of antennas is often a strong function of frequency.  These two effects can impart frequency structure on the intrinsically smooth foregrounds.  Synchrotron foregrounds are often about 10\% polarized and are 1000 times brighter than the HI signal at low redshift, so ideally one would like to minimize polarization leakage so that leakage from this large polarized signal does not overwhelm the 21~cm signal.
The method for restricting polarization leakage is clear from equations \ref{eq:t}, \ref{eq:l}, and \ref{eq:I}.  To minimize the angle dependent cross terms in equations \ref{eq:t} and \ref{eq:l}, the cross-polar response should be much smaller than the co-polar response.  When this is true, only the first term is significant in equations \ref{eq:t} and \ref{eq:l}.  Figure \ref{fig:crosspol} demonstrates that we have achieved this for the Tianlai beam: the cross-polar response is much lower than the co-polar response in the focusing and non-focusing cuts for both polarizations.  Polarization leakage can then be fully mitigated if the co-polar response of the transverse and longitudinal polarizations are equal.  This is extremely difficult to achieve, and a glance at figure \ref{fig:crosspol} shows that the co-polar responses of the two polarizations are indeed not equal.  In practice, some degree of polarization leakage is unavoidable, and the polarized m-mode formalism of \cite{shaw2015coaxing} demonstrates that, by restricting the analysis to m-modes with minimal leakage, unbiased 21 cm maps can still be made, provided that one can model the polarized beam patterns accurately.

\section{Conclusions} \label{concl}
A new `four-hex' feed antenna for radio interferometer arrays has been designed, built, and tested. The feed has more than an octave of bandwidth, excellent isolation between polarizations, and measured beam patterns that are in good agreement with computer simulations. As expected, when the feeds are placed in a linear array, coupling between feeds causes broadening of the beam patterns. These simulations and measurements will be used to simulate the visibilities that will be measured by the Tianlai cylindrical radio telescope.  The noise and sky coverage properties meet or exceed those assumed in the forecast of \cite{Xu2015}.  Successful foreground removal will depend on calibrations to accurately measure the true beam pattern.

\section*{Acknowledgments}
We gratefully acknowledge the help of Mr. Ting-Yen Shih, who performed the feed antenna pattern measurements, and Prof. Nader Behdad, who made his antenna test range available to us at the UW-Madison Department of Electrical and Computer Engineering. Work at UW-Madison was partially supported by NSF Award AST-1211781. Work at NAOC was supported by the MOST 863 program 2012AA121701, the NSFC 11633004, the CAS QYZDJ-SSW-SLH017 and the CAS Strategic Priority Research program XDB09020301.

\bibliographystyle{abbrv}
\bibliography{bibliography}

\end{document}